\documentclass[10pt, a4paper]{article}
\usepackage{color}
\usepackage{amssymb,bm}%símbolo de conjuntos
\usepackage{graphicx}%figuras
\usepackage{amsthm}%definir comandos newtheorem
\usepackage{stmaryrd}%setas
\usepackage{amsmath}%símbolos matemáticos

\newtheorem{teo}{Theorem}
\newtheorem{lem}{Lemma}
\usepackage[utf8]{inputenc}
\usepackage[T1]{fontenc}
\usepackage{tikz}
\usepackage[left=2cm,right=1.5cm,top=2cm,bottom=1.5cm]{geometry}
\usepackage{verbatim}
\author{Paulo C. Lima\footnote{Departamento de Matem{\'a}tica, Universidade Federal de Minas Gerais, Av. Ant\^onio
		Carlos 6627 C.P. 702 CEP 30123-970 Belo Horizonte-MG, Brazil} \and Ricardo Lopes de Jesus$^*$  \and Aldo Procacci$^*$ }
\title{Absolute convergence of the free energy of the BEG model in the disordered region for all temperatures}
\date{}

\begin{document}
\maketitle

\begin{abstract} We analyze the $d$-dimensional  Blume-Emery-Griffiths model in the disordered region of parameters and we show
that its free energy can be  explicitly written in term of a  series which is absolutely convergent at any temperature
in an unbounded portion of this region. As a byproduct we also obtain an upper bound for the number of $d$-dimensional fixed polycubes
of size $n$.
\end{abstract}

\let\a=\alpha \let\b=\beta \let\ch=\chi \let\d=\delta \let\e=\varepsilon
\let\f=\varphi \let\g=\gamma \let\h=\eta    \let\k=\kappa \let\l=\lambda
\let\m=\mu \let\n=\nu \let\o=\omega    \let\p=\pi \let\ph=\varphi
\let\r=\rho \let\s=\sigma \let\t=\tau \let\th=\vartheta
\let\y=\upsilon \let\x=\xi \let\z=\zeta
\let\D=\Delta \let\F=\Phi \let\G=\Gamma \let\L=\Lmbda \let\Th=\Theta
\let\O=\Omega \let\P=\Pi \let\Ps=\Psi \let\Si=\Sigma \let\X=\Xi
\let\Y=\Upsilon\let\L\Lambda
\def\Zd{{\mathbb{Z}^d}}

\section{The model. Notations and results}

\def\x{{\mbox{\footnotesize{X}}}}
\def\y {{\mbox{\footnotesize{\rm Y}}}}
\def\xx{{\mbox{\scriptsize{\rm X}}}}
\def\yy {{\mbox{\scriptsize{Y}}}}
\def\ss{{\bm\sigma}}
\def\ni{\noindent}
\def\GG{{\mathcal G}}
\def\PP{{\mathcal P}}
\def\0{\emptyset}
\def\be{\begin{equation}}
\def\ee{\end{equation}}

The Blume-Emery-Griffiths (BEG) model is a spin-one system, introduced in the 1970s in order to explain some of the physical properties of  $\,{}^3\mathrm{He}-{}^4\mathrm{He}$ mixtures \cite{bib:rbeg} and since then  it has attracted a lot of attention and has been used in several  applications such as ternary fluids \cite{bib:muk,bib:fur}, phase transitions in $UO_2$ \cite{bib:grif} and $DyVO_4$ \cite{bib:blume} and phase changes in microemulsion \cite{bib:ss}.

\ni
The BEG model is defined  in  the $d$-dimensional cubic lattice $\Zd$   by
supposing that in each site $x\in \Zd$ there is a random  variable $\s_x$ (the spin at $x$) taking values in the
set $\{0,\pm 1\}$.
For $U\subset \Zd$, a spin configuration $\bm\s_U$  in  $U$ is a function
$\bm\s_U: U\to \{0,\pm 1\}: x\mapsto \s_x$ and $\Si_U$ will denote the set of all spin configurations in $U$.
Given a finite set $\L\subset \Zd$ (typically a cubic box centered at the origin of $\Zd$),
the Hamiltonian of the system in $\L$ (with zero boundary conditions and  zero magnetic field) has the following expression:
\begin{align}\label{hamiltoniana1}
{\mathcal H}_\L(\bm\sigma_\L)=-\sum_{\{x,y\}\subset \L} (\sigma_x\sigma_y+\y\sigma_x^2\sigma_y^2)\d_{|x-y|1}-2d\,\x\sum_{x\in \L}\sigma_x^2,
\end{align}
where $|\cdot|$ is the usual $L^1$ norm in $\Zd$, $\d_{|x-y|1}$ is the Kronecker symbol (i.e. $\d_{|x-y|1}=1$ if $|x-y|=1$ and zero otherwise) and $\x,\y\in\mathbb{R}$.

\ni
The parameter space $\x\y$ of the model is generally partitioned  in three distinct regions (ferromagnetic, disorderd, anti-quadrupolar) according to the ground state configurations of the Hamiltonian (see for instance \cite{bib:eumg}). In this note, we will focus our attention in the
disordered region, namely  $${\mathcal D}=\{(\x,\y)\in\mathbb{R}^2:\x<0, \, 1+2\x+\y<0\},$$  where the unique ground state is the constant configuration $\sigma_x=0$, for all $x\in \L$.

\ni
The probability $P_\L(\ss_\L)$ (i.e. the finite volume Gibbs measure) of a configuration $\ss_\L\in \Sigma_\L$ is defined as
$$
P_\L(\ss) =\frac{ e^{-\b {\mathcal H}_\L(\ss_\L)}}{Z_\Lambda(\x,\y,\beta)}
$$
where  $\beta$ the inverse of the temperature in units of the  Boltzmann constant and the normalization constant
$Z_\Lambda(\x,\y,\beta)$ is the   partition function of the  model  given by
\begin{equation}\label{part}
Z_\Lambda(\x,\y,\beta)=\sum_{\ss_\L\in \Sigma_\Lambda} e^{-\beta H_\Lambda(\ss)},
\end{equation}
Finally, the free energy (in fact, the pressure) in the thermodynamic limit is defined as

\begin{equation}\label{energia}
f(\x,\y,\beta)=\lim_{\Lambda\uparrow \mathbb{Z}^d}\frac{1}{|\Lambda|}\log Z_\Lambda(\x,\y,\beta),
\end{equation}
where  $|\Lambda|$ is the cardinality of $\Lambda$.  The  limit in the r.h.s. of (\ref{energia}) is taken, for instance,  in van Hove sense, is well defined, it is  independent of the sequence $\Lambda\uparrow \mathbb{Z}^d$ and on the boundary condition, see for instance \cite{bib:sv}.

\ni
In \cite{MP}
it is shown that for a certain class of spin systems interacting via a pair potential (which includes the BEG model) the free energy
can be written in terms of  a series which is absolutely convergent  at  any temperature if some conditions on the pair potential are satisfied.
Concerning specifically  the BEG model, in  \cite{bib:eu2018} it is proved that there exists   a region of parameters  ${\cal D}_{Dob}\subset {\cal D}$ where the Dobrushin Uniqueness Criterium is satisfied for all temperature.
In the present note, using the scheme described in \cite{MP} and a recent tree graph inequality proved in \cite{PY}, we establish an optimal  region  ${\cal D}_{analytic}\subset {\cal D}_{Dob}$ where the free  energy of the BEG model can be explicitly written as an absolutely  convergent series of analytic functions for all temperature, ruling out the presence of phase transitions of any order in this region.
Our main result can be resumed by the following theorem.

\begin{figure}[h!]
	\centering
	\begin{tikzpicture}[scale=0.5]
	\shade[top color=gray] (-11.812,-6.87)--(-3.002,-1)--(-5.806,-1)--(-2.953,0)--(-11.812,3)--cycle;
	\draw[->] (-2.953,0) -- (2,0) node[right] {$\x$};
	\draw[->] (0,-7) -- (0,4) node[above] {$\y$};
	\draw (-4,4) node[below left] {$\x+k\y+k=0$};
	\draw [<-](-8,1.8) .. controls (-8,2) and (-7.5,2.5) .. (-7.5,3);
	\draw (-2,-4) node[below left] {$\x-\bar{k}\y+\bar{k}=0$};
	\draw [<-](-5.8,-3) .. controls (-5.3,-3.5) and (-5.5,-3.5) .. (-5,-4);
	\draw (-6.6,2) node[below right] {$\x-(k-1)\y+k=0$};
	\draw [<-](-3.8,-0.4) .. controls (-1,-1) and (-1,0.5) .. (-1,1);
	\draw[domain=-3.002:0,color=black,dashed] plot (\x,-1);
	\draw (0,-0.5) node[below right] {$-1$};
    \draw (-10,-0.5) node[below right] {${\cal D}_{analytic}$};
	\end{tikzpicture}
	\caption{ The region of analyticity of the free energy for all temperature.}
	\label{fig1}
	\end{figure}
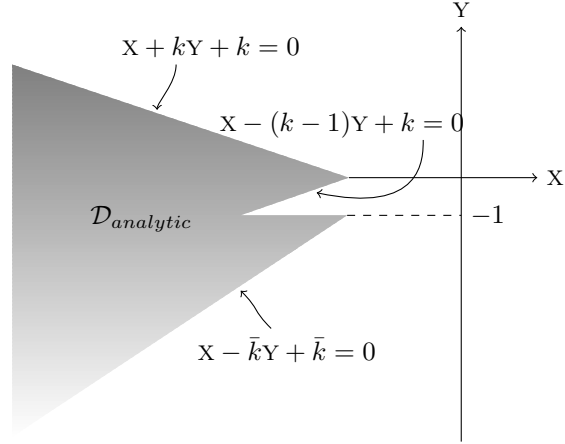

\begin{teo}\label{teo1} Let $d\geq 2$ and  let ${\cal D}_{analytic}\subset {\cal D}$ be the region whose boundary is the polygonal curve
 \begin{eqnarray}\x=\left\{\begin{array}{ll}
 -k(\y+1),& \mbox{if $\y\geq 0$}\\
 (k-1)\y-k,&\mbox{if $-1<\y< 0$}\\
 \bar{k}(\y-1),&\mbox{if $\y\leq -1$}
 \end{array} \right.,
 \end{eqnarray}
  where $k={59.56d-1\over 2d}$ and $\bar{k}={30.52d-1\over 2d}$  (see Figure \ref{fig1}).  If $(\x,\y)\in {\cal D}_{analytic}$, then the free energy of the BEG model defined in (\ref{energia})  can be explicitly written as an absolutely  convergent series of analytic functions at any temperature.
\end{teo}
\ni
%%It is worth to stressing that in  \cite{bib:eu2018} one of us   obtained a region of parameters  ${\cal U}_{Dob}\subset {\cal D}$, where the Dobrushin Uniqueness Criterium is satisfied for all temperature, showing the uniqueness of the  (infinite volume) Gibbs measure for all temperature.
%The uniqueness of the Gibbs measure assures the absence of the first order phase transition; however, it does not prevent the possibility of having high-order phase transitions.
% One way to assure this is not the case, would be to analyze the analyticity properties the free energy.
%The analyticity  of the free energy and correlations for a large class of spin systems at high temperature or small fugacity have been inferred via the  Dobrushin Uniqueness Criterium, in \cite{I}.
%In particular by checking condition (2) in Lemma II.2  of \cite{I} for the BEG model (which, if veriefied, implies the  Dobrushin uniqueness condition), the one can use  Theorem II.3 in \cite{I} to conclude on analyticity.

\ni
The rest of this note is organized as follows. In section \ref{secproof} we prove Theorem \ref{teo1} and in Section \ref{poly} we
use the technique developed in Section \ref{secproof} (in particular Lemma \ref{lemma2}) to obtain
 an upper bound on the  number $A_n$ of $d$-dimensionl polycubes of size $n$  (for the definition of  a  $d$-dimensionl polycube we refer the reader to  Section \ref{poly}).

%\section{Notation and main result} \label{secnot}

\section{Proof of Theorem  \ref{teo1}}\label{secproof}
\ni
We start by rewriting the partition function of the model defined in (\ref{part}) performing a so-called  high temperature polymer expansion.
From   ~\eqref{hamiltoniana1} and  ~\eqref{part}, we have

\begin{equation}\label{parti}
Z_\Lambda(\x,\y,\beta)=\sum_{\ss_\L\in \Sigma_\Lambda} e^{ 2d\beta\, \xx\sum_{x\in \Lambda}\sigma_x^2}e^{\beta \sum_{\{x,y\}\subset \L} (\sigma_x\sigma_y+\yy\sigma_x^2\sigma_y^2)\d_{|x-y|1}}.
\end{equation}
By Mayer expansion  the second exponential in the r.h.s. of  ~\eqref{parti} can be  written as
\begin{eqnarray*}\label{exp}
e^{\beta  \sum_{\{x,y\}\subset \L} (\sigma_x\sigma_y+\y\sigma_x^2\sigma_y^2)\d_{|x-y|1}} &=&\prod_{\{x,y\}\subset \Lambda}
[e^{\beta(\sigma_x\sigma_y+\yy\sigma_x^2\sigma_y^2)\d_{|x-y|1}}-1+1]\\
&=&\sum_{g\in \mathcal{G}_\Lambda} \prod_{\{x,y\} \in E_{g}} (e^{\beta (\sigma_x\sigma_y+\yy\sigma_x^2\sigma_y^2)\d_{|x-y|1}}-1)\\
%&=&\sum_{g\in \mathcal{G}_\Lambda} \prod_{\{x,y\} \in E_{g}} (e^{\beta (\sigma_x\sigma_y+\yy\sigma_x^2\sigma_y^2)}-1)\d_{|x-y|1}
\end{eqnarray*}
where  $\mathcal{{G}}_\Lambda$ is the set of all possible graphs (connect or not) with vertex set  $\Lambda$ and given $g\in \mathcal{{G}}_\Lambda$, $E_g$ denotes its edge set.
Let  $\Pi_n(\Lambda)$ be the set of all partitions of $\Lambda$ having  $n$ elements ($n=1,2, \dots, |\L|$), then, denoting shortly
$$
F_{xy}= e^{\beta (\sigma_x\sigma_y+\yy\sigma_x^2\sigma_y^2)\d_{|x-y|1}}-1,
$$
it is a standard observation in the framework of the cluster expansion techniques (see for instance \cite{Ca, bib:sv}), that

\begin{equation*}
\sum_{g\in \mathcal{G}_\Lambda}\prod_{\{x,y\}\in E_g}F_{xy} = \sum_{n=1}^{|\L|}\sum_{\{R_1,...,R_n\}\in \Pi_n(\Lambda)} \prod_{l=1}^{n} \left(\sum_{g\in G_{R_l}}\prod_{\{x,y\}\in E_g}F_{xy}\right).
\end{equation*}
where in the r.h.s. $G_{R_l}$ denotes the set of all connecetd graphs with vertex set $R_l$.
So, we have

\begin{equation*}
e^{\beta \sum_{\{x,y\}\subset \L} (\sigma_x\sigma_y+\yy\sigma_x^2\sigma_y^2)\d_{|x-y|1}}=
\sum_{n=1}^{|\L|}\sum_{\{R_1,...,R_n\}\in \Pi_s(\Lambda)} \prod_{i=1}^{n} \r(R_i, \s_{R_i})
\end{equation*}
where, for  $R\subset \L$
\begin{displaymath}
\rho(R,\s_{R})=\left\{ \begin{array}{cc}
1  & \mbox{if $|R|=1$}\\ \\
\sum\limits_{g\in G_{R}}\prod\limits_{\{x,y\} \in E_{g}}(e^{\beta\a (\sigma_x\sigma_y+\yy\sigma_x^2\sigma_y^2)\d_{|x-y|1}}-1)  & \mbox{if $|R|\geq 2$}
\end{array}\right..
\end{displaymath}

\ni
We have therefore the following representation for the partition function  ~\eqref{parti}

\begin{eqnarray}\label{fpart}
Z_\Lambda(\x,\y,\beta)&=& \sum_{\ss_\L\in \Sigma_\Lambda} e^{ 2d\beta\, \xx\sum_{x\in \Lambda}\sigma_x^2}\sum_{n=1}^{|\L|}\sum_{\{R_1,...,R_n\}\in \Pi_n(\Lambda)} \prod_{l=1}^{n} \rho(R_l,\ss_{R_l})\nonumber \\
&=& \sum_{\ss_\L\in \Sigma_\Lambda} \sum_{n=1}^{|\L|}\sum_{\{R_1,...,R_n\}\in \Pi_n(\Lambda)} \prod_{l=1}^{n} e^{ 2d\beta\, \xx\sum_{x\in R_l}\sigma_x^2} \rho(R_l,\ss_{R_l}) \nonumber \\
&=& \sum_{n=1}^{|\L|}\sum_{\{R_1,...,R_n\}\in \Pi_n(\Lambda)}\prod_{l=1}^n\Big( \sum_{\ss_{R_l}\in \Sigma_{R_l}}  \rho(R_l,\ss_{R_l}) e^{ 2d\beta\, \xx\sum_{x\in R_l}\sigma_x^2}\Big),
\end{eqnarray}
Defining $$\bar{\rho}(R)=\sum_{\ss_R\in \Sigma_{R}}  \rho(R, \ss_R) e^{ 2d\beta\, \xx\sum_{x\in R}\sigma_x^2},$$
the  partition function  ~\eqref{fpart} can be rewritten as

\begin{equation*}
Z_\Lambda(\x,\y,\beta)=\sum_{n=1}^{|\L|}\sum_{\{R_1,...,R_n\}\in \Pi_n(\Lambda)}\bar{\rho}(R_1)\cdots\bar{\rho}(R_n).
\end{equation*}
Notice that, if $|R|=1$ and hence $R=\{x\}$ with $x\in \L$,  we have
$$\bar{\rho}(R)=\bar\r(\{x\})= \sum_{\s_x\in \{0,\pm1\}} e^{ 2d\beta\, \xx\sigma_x^2}=1+2e^{2d\beta\, \xx}.$$
Moreover  observe that for any $R\subset \mathbb{Z}^d$ such that $|R|\ge 2$  and  for any  $g\in G_R$  the factor
$$\prod_{\{x,y\} \in E_{g}} (e^{\beta (\sigma_x\sigma_y+\yy\sigma_x^2\sigma_y^2)\d_{|x-y|1}}-1)$$ is equal to zero whenever
$\sigma_x =0$ for some $x\in R$. Thus, defining
$$
\bar{\Sigma}_R=\{\ss_R\in \Sigma_R : \sigma_x=\pm 1, \ \forall\ x\in R\},
$$
we can write, for  $|R|\geq 2$
\begin{eqnarray}\label{eqpxx} \bar{\rho}(R)=\sum_{\sigma_{R}\in \bar{\Sigma}_{R}}  \rho(R,\ss_R) e^{ 2d\beta\, \xx\sum_{x\in R}\sigma_x^2}=\sum_{\sigma_{R}\in \bar{\Sigma}_{R}}  \rho(R,\ss_R) e^{ 2d\beta\, \xx|R|}.\end{eqnarray}
Hence setting
\begin{equation}\label{atividade}
\xi(R)=\left\{\begin{array}{cc}
1 & \mbox{if $|R|=1$} \\ \\
\left(\frac{e^{2d\beta\,\xx}}{1+2e^{2d\beta\, \xx}}\right)^{|R|}\sum\limits_{\sigma_{R}\in \bar{\Sigma}_{R}}\limits\sum\limits_{g\in G_{R}}\prod\limits_{\{x,y\} \in E_{g}} (e^{\beta (\sigma_x\sigma_y+\y)\d_{|x-y|1}}-1) & \mbox{ if $|R|\geq 2$}
\end{array}\right.,
\end{equation}
we can rewrite the partition function as
\begin{eqnarray*}
Z_\Lambda(\x,\y,\beta)&=&(1+2e^{2d\beta\, \xx})^{|\Lambda|}\sum_{n=1}^{|\L|}\sum_{\{R_1,...,R_n\}\in \Pi_n(\Lambda)}\xi(R_1)\cdots\xi(R_n) \nonumber \\
&=& (1+2e^{2d\beta \x})^{|\Lambda|}\left(1+\sum_{n\ge 1}\sum_{{\{R_1,...,R_n\}:\;R_i\subset \Lambda \atop |R_i|\geq 2,R_i\cap R_j= \emptyset}}\xi(R_1)\cdots\xi(R_n)\right)
\nonumber \\
&=& (1+2e^{2d\beta \x})^{|\Lambda|}\left(1+\sum_{n\geq 1}\frac{1}{n!}\sum_{(R_1,...,R_n)\in \Lambda^n\atop |R_i|\geq 2,R_i\cap R_j= \emptyset}\xi(R_1)\cdots\xi(R_n)\right) \nonumber \\
&\equiv&(1+2e^{2d\beta \x})^{|\Lambda|} \ \ \Xi_\Lambda(\x,\y,\beta),
\end{eqnarray*}
where $\Xi_\Lambda(\x,\y,\beta)$ is the  grand canonical   partition function of an abstract  polymer model, in which the polymer  $R$ is a finite subset of $\Zd$ with cardinality greater than $1$, with activity $\xi(R)$ given by (\ref{atividade}) and with incompatibility relation being the non-empty intersection  (see e.g. \cite{KP, FP, BFP}).

\ni
Therefore, the free energy of the system  (in the finite volume  $\Lambda$) is given by  $$f_\Lambda(\x,\y,\beta)=\frac{1}{|\Lambda|}\log Z_\Lambda(\x,\y,\beta)= \log(1+2e^{2d\beta\, \x})+P_\Lambda(\x,\y,\beta),$$
where
\begin{equation*}\label{pressao}
P_\Lambda(\x,\y,\beta)=\frac{1}{|\Lambda|}\log \Xi_\Lambda(\x,\y,\beta).
\end{equation*}
Since $\log(1+2e^{2d\beta \xx})$ is analytic for all $\beta\in \mathbb{R}$, it is enough to study the absolute convergence of  the pressure $P_\Lambda (\xx,\yy,\beta)$ of the polymer gas described above  as a function of $\b$ in order to get information about analyticity of the free energy in the thermodynamic limit.

\ni
The conditions for  the
absolute convergence and boundness (uniformly in $\L$) of the pressure of an abstract polymer gas such as the one described above have been studied since a long time (see for instance \cite{BFP} and references therein). We apply here the  Fernandez-Procacci (FP) criterium \cite{FP}, according to which the pressure $P_\Lambda (\xx,\yy,\beta)$ can be written as series which converges absolute  uniformly bounded in the volume $\Lambda$, as long as the condition below is satisfied

\begin{equation*}
\sum_{n\geq 2} e^{a n}\sup_{z\in \mathbb{Z}^d} \sum_{\overset{R \subset \mathbb{Z}^d : z \in R}{|R|=n}}|\xi(R)|\leq e^a-1.
\end{equation*}
where $\xi(R)$ is defined in (\ref{atividade}) and $a>0$ is an arbitrary parameter to be optimized. Choosing $a=\log 2$  (which is not far from the optimal value) and using translational invariance, this condition becomes

\begin{equation}\label{criterio}
\sum_{n\geq 2} 2^{ n}\sum_{R \subset \mathbb{Z}^d \atop 0 \in R, |R|=n}|\xi(R)|\leq 1.
\end{equation}
where $0\in R$ means that $R$ contains the origin of $\Zd$.
Setting
\begin{equation}\label{alpha}
\alpha(\x,\beta)=\frac{e^{2d\beta \,\x}}{1+2e^{2d\beta \,\x}},
\end{equation}
we have
\begin{equation}\label{sup1}
\sum_{R \subset \mathbb{Z}^d \atop 0 \in R, |R|=n}|\xi(R)|\leq [\alpha(\x,\beta)]^n \sum_{R \subset \mathbb{Z}^d \atop 0 \in R, |R|=n}\sum_{\ss_{R}\in \bar{\Sigma}_{R}}
\left|\sum_{g\in G_{R}}\prod_{\{x,y\} \in E_{g}} (e^{\beta (\sigma_x\sigma_x+\yy)\d_{|x-y|1}}-1)\right|.
\end{equation}
We now  need an upper bound for the  factor
\begin{equation}\label{fact}
\left|\sum_{g\in G_{R}}\prod_{\{x,y\} \in E_{g}} (e^{-\beta V_{xy}}-1)\right|
\end{equation}
where we have denoted shortly
\begin{equation}\label{vxy}
V_{xy}=-(\sigma_x\sigma_y+\y)\d_{|x-y|1}
\end{equation}
Let us first remark  that  the pair potential $V_{xy}$ defined in (\ref{vxy}) enjoys the so-called stability property   (see e.g. \cite{Ru}) accordingly to the  following lemma.
\begin{lem}
For any $R\subset \Zd$ and for any $\ss_\L\in \Sigma_\L$ it holds that
\begin{equation}\label{sta}
\sum_{\{x,y\}\subset R}V_{xy}\ge - h(\y)|R|
\end{equation}
where
\begin{equation}\label{csta}
h(\y)=\begin{cases} d(1+\y) & if ~\y>-1\\
0 & if ~\y\le -1
\end{cases}.
\end{equation}
\end{lem}
\ni
{\it Proof}.  Consider first the case $\y>-1$. Observe that   $\s_x\s_y\ge -1$ and therefore $V_{xy}\geq -(1+\yy)\d_{|x-y|1}$. Moreover for  each
$R\subset \L$ and $x\in R$ we have that $\sum_{y\in R\atop y\neq x}\d_{|x-y|1}\le 2d$. Hence, given $R\subset \Zd$ and $\ss_R\in \Sigma_R$, we can bound

$$
\sum_{\{x,y\}\subset R}V_{xy}
=
{1\over 2}\sum_{(x,y)\in R^2\atop y\neq x}V_{xy}= {1\over 2} \sum_{x\in  R}\sum_{y\in R\atop y\neq x}V_{xy}
\ge
-{(1+\y)\over 2}\sum_{x\in  R}\sum_{y\in R\atop y\neq x}\d_{|x-y|1} \ge -d(1+\y)|R|
$$
The case $\y\leq -1$ is trivial.  Indeed, when $\y\le-1$ we  have  $V_{xy}\geq 0$ for all $\ss_{\{x,y\}}\in \Sigma_{\{x,y\}}$ and therefore, for any $R$ and any $\ss_R\in \Sigma_R$
$$
\sum_{\{x,y\}\subset R}V_{xy}\ge 0
$$
This concludes the proof of the Lemma. $\Box $

\vskip.15cm

\ni
Let us now go back to the problem of finding an upper bound for (\ref{fact}). As long as the pair potential
$V_{xy}$ is stable (i.e. satisfies (\ref{sta})), it is long known that
efficient  estimates on  factors of the form (\ref{fact}) involving sum over connected graphs  can be obtained via tree graph identities and tree graph inequalities.
Here we will use a recent tree graph inequality due to Procacci and Yuhjtman (Proposition 1 in \cite{PY}). By such an inequality we can bound in our present case, for any $R\subset \Zd$ such that
$|R|=n$ and any $\ss_R\in \Sigma_R$,

\begin{eqnarray}\label{modulo}
\left|\sum_{g\in G_{R}}\prod_{\{i,j\} \in E_{g}} (e^{-\beta V_{xy}}-1)\right| &\leq & e^{\beta n h(\yy)}\sum_{\tau \in T_{R}}\prod_{\{x,y\} \in E_{\tau}}(1-e^{-\beta |V_{xy}|})\nonumber \\
&\leq& e^{\beta n h(\yy)}\sum_{\tau \in T_{R}}\prod_{\{x,y\} \in E_{\tau}}(1-e^{-\beta (1+|\y|)\delta_{|x-y|1}})\nonumber \\
&=& e^{\beta n h(\yy)}\sum_{\tau \in T^*_{R}}\prod_{\{x,y\} \in E_{\tau}}(1-e^{-\beta (1+|\y|)\delta_{|x-y|1}}).
\end{eqnarray}
where $T_R$ denotes the set of all tree graphs   with vertex set $R$ (i.e. connected graphs without loops) and $T^*_R$ is the subset of $T_R$ formed by all trees with maximum degree less than or equal to $2d$.

\ni
Turning back now to (\ref{sup1}), first observe  that the right hand side of the inequality (\ref{modulo}) has no dependence on the spins and thus  we can freely sum in the r.h.s. of (\ref{modulo}) over all configurations in $\bar \Sigma_{R}$ getting  a factor $2^{|R|}=2^n$. Secondly, notice that
$$
1-e^{-\beta (1+|\yy|)\delta_{|x-y|1}}= (1-e^{-\beta (1+|\yy|)})\delta_{|x-y|1}
$$
implies that, for any tree $\t\in T_R^*$
$$
\prod_{\{x,y\} \in E_{\tau}}(1-e^{-\beta (1+|\yy|)\delta_{|x-y|1}})= (1-e^{-\beta (1+|\yy|)})^{|R|-1} \prod_{\{x,y\} \in E_{\tau}}\delta_{|x-y|1}
$$
These two facts, together with ~\eqref{modulo} implies that ~\eqref{sup1} can be written as
\begin{equation}\label{sup2}
\sum_{R \subset \mathbb{Z}^d \atop  0 \in R,\, |R|=n}|\xi(R)|\leq [2 \alpha(\x,\b) e^{\beta h(\yy)} ]^n
(1-e^{-\beta (1+|\yy|)})^{n-1}\sum_{R \subset \mathbb{Z}^d \atop  0 \in R,\, |R|=n}\sum_{\tau \in T^*_{R}}\prod_{\{x,y\} \in E_{\tau}}\delta_{|x-y|1}.
\end{equation}
Set now
\be\label{Cen}
C_n=\sum_{R \subset \mathbb{Z}^d \atop  0 \in R,\, |R|=n}\sum_{\tau \in T^*_{R}}\prod_{\{x,y\} \in E_{\tau}}\delta_{|x-y|1}
\ee
and notice that
$$\sum_{R \subset \mathbb{Z}^d \atop  0 \in R,\, |R|=n}(\ldots)=\sum_{\{x_1,...,x_n\} \subset \mathbb{Z}^d\atop  x_1=0}(\ldots)= \frac{1}{(n-1)!}\sum_{(x_1,...,x_n) \in (\mathbb{Z}^{d})^n\atop x_1=0,~ x_i\neq x_j}(\ldots),
$$
So, letting  $T^*_n$ denote the set of all trees  with vertex set $[n]$ and degrees  $d_i$ such that $1\le d_i\le 2d$ , we have
$$
C_n= \frac{1}{(n-1)!}\sum_{\t\in T^*_n}\sum_{(x_1,...,x_n) \in (\mathbb{Z}^{d})^n\atop x_i\neq x_j,\; x_1=0}\prod_{\{i,j\} \in E_{\tau}}\delta_{|x_i-x_j|1}
=~{1\over (n-1)!}\sum_{\t\in T^*_n}
w_\t
$$
where we have set
\be\label{wtau0}
w_\t~= ~\sum_{(x_1,\dots, x_n)\in \mathbb{Z}^{dn}:\atop x_1=0,~x_i\neq x_j}
\prod_{\{i,j\}\in E_\t}\d_{|x_i-x_j|1}
\ee
The following lemma provides an upper bound for $w_\t$.
\begin{lem}\label{lemma2}
Given a tree $\t\in T^*_n$ with degree $d_1,\dots d_n$ at vertices $1,\dots ,n$, we have
\be\label{bwtau}
w_\t \,\le\,
{2d!\over (2d-d_1)!} \prod_{i=2}^{n} {(2d-1)!\over (2d-d_i)!}
\ee
\end{lem}

\ni
{\it Proof}. We will consider $\t$ as rooted in  $1$. We recall that a vertex $j\neq 1$ of $\t$ such that $d_j=1$ is called a leaf of $\t$.  Moreover, given a vertex $j$ of $\t$ with degree $d_j$,
we denote by  $j'$  the(unique)  vertex of $\t$ which is  the father of $j$ in $\t$ and we denote by  $j_1,\dots,j_{d_j-1}$  the vertices of $\t$ which are  the  children of $j$.
Let $I=\{i\in [n]: \; i>1~{\rm and}~d_i>1\}$. Let $J$ be the subset of $I$ formed by the vertices of $\t$ whose
children are all leaves.
For each $j\in J$,  we can perform  the sum over $x_{j_1},\dots,x_{j_{d_j-1}}$ and we get
$$
\sum_{(x_{j_1},\dots,x_{j_{d_j-1}})\in (\Zd) ^{d_j-1}\atop  x_{j_k}\neq x_{j'},~x_{j_i}\neq x_{j_k}}\prod_{s=1}^{d_j-1} \d_{|x_j-x_{j_s}|1}= (2d-1)(2d-2)\dots (2d-(d_j-1))={(2d-1)!\over (2d-d_j)!}
$$
Then the sum over all coordinates $\{x_{j_1},\dots x_{j_{d_j-1}}\}_{j\in J}$ produce (at most) a factor
$$
\prod_{j\in J}  {(2d-1)!\over (2d-d_j)!}
$$
and we are left with a ``defoliated" tree $\t'\subset \t$ such that now all $j\in J$ are leaves.
Iterating this procedure,  observing that when  $j=1$  the sum over the coordinate its children (when  they are all leaves) produces the factor
$$
2d(2d-1)\cdots (2d-d_1+1)= {(2d)!\over (2d-d_1)!}
$$
we get
$$
w_t\le   {(2d)!\over (2d-d_1)!}\prod_{j\in I}{(2d-1)!\over (2d-d_j)!}
$$
where the inequality is due to the fact that  we are not taking into account  that we could produce cycles when we sum over  the coordinates
$\{x_j\}_{j\in I}$.
Now  observing that when $j\notin I$ (i.e. $d_j=1$ and $j\neq 1$) we have that
$$
{(2d-1)!\over (2d-d_j)!}=1
$$
inequality (\ref{bwtau}) follows. $\Box$
\vskip.2cm
\ni
Inequality (\ref{bwtau}) will be used in the next section to obtain an upper bound on the number of $d$-dimensional fixed polycubes
of size $n$. For the purpose of this section it is sufficient to use a simplified (and slightly  worse) estimate for $w_\t$ easily derived from (\ref{bwtau}).
Indeed, from (\ref{bwtau}) we have that
\begin{eqnarray}\label{wtau}
w_\t &\le&
{2d!\over (2d-d_1)!} \prod_{i=2}^{n} {(2d-1)!\over (2d-d_i)!}\\
&\le &  2d\prod_{i=1}^n (2d-1)^{d_i-1}= 2d (2d-1)^{n-2}\nonumber
\end{eqnarray}
where in the last inequality we have used that for any tree $\t\in T_n$ it holds that $\sum_{i=1}^n d_i = 2n-2$. Therefore we can bound
\begin{eqnarray}
C_n & \le & \;{2d\over (n-1)!} (2d-1)^{n-2} \sum_{\t\in T^*_n}1\nonumber \\
& \le &\; {2d\over (n-1)!} (2d-1)^{n-2} \sum_{\t\in T^*_n}1\nonumber \\
& = & 2d {n^{n-2}\over (n-1)!} (2d-1)^{n-2}\nonumber \\
&\le &   {n^{n-2}\over (n-1)!} (2d)^{n-1}
\end{eqnarray}
where in the third line we have used   Cayley formula (which says that  $\sum_{\tau \in T_{n}}1=n^{n-2}$).

\ni
In conclusion
 l.h.s. of \eqref{sup2} is bounded   as
\begin{eqnarray*}\label{sup3}
 \sum_{R \subset \mathbb{Z}^d :\, 0\in R\atop |R|=n}|\xi(R)|&\leq&  [2 \alpha(\x,\b) e^{\beta h(\yy)}]^n
\left[2d(1-e^{-\beta (1+|\yy|)})\right]^{n-1}\frac{n^{n-2}}{(n-1)!}
\end{eqnarray*}
By  Stirling formula, we get the inequality $\frac{n^{n-2}}{(n-1)!}\leq \frac{e^{n-1}}{n}$, which used  in the above expression, implies that  condition  \eqref{criterio} is satisfied  provided that
\begin{equation*}
	\sum_{n=2}^{\infty}[4 \alpha(\x,\b) e^{\beta h(\yy)}]^n \frac{e^{n-1}}{n}\left[2d(1-e^{-\beta (1+|\yy|)})\right]^{n-1}\leq 1,
\end{equation*}
or

\begin{equation}\label{alltemp}
4 \alpha(\x,\b) e^{\beta h(\yy)} \sum_{n=2}^{\infty} {1\over{n}}\left[8de \alpha(\x,\b)  e^{\beta h(\yy)} (1-e^{-\beta (1+|\yy|)})\right]^{n-1}\leq 1.
\end{equation}
When  $\beta=0$, the condition (\ref{alltemp}) is trivially satisfied. Let us thus suppose $\b>0$ and let
$\delta=\alpha(\x,\b)  e^{\beta h(\yy)}$ and  $\epsilon=(1-e^{-\beta (1+|\yy|)})$. Using the fact that  $\sum_{n=2}^\infty {r^{n-1}\over n} = {-r-\log(1-r) \over r}$ for $|r|<1$, the inequality above  will be satisfied, as long as

\begin{eqnarray*}
4\delta\,\,\frac{[-8de\delta \epsilon-\log(1-8de\delta\epsilon)]}{8de\delta \epsilon}\leq 1,
\end{eqnarray*}
or, equivalently,

\begin{eqnarray}\label{eqcondx}
4\delta\left[-1-\log{(1-8de\delta\epsilon)}^{1\over 8de\delta\epsilon}\right]\leq 1.
\end{eqnarray}
Since $\delta$ is a function of  the stability constant $h(\y)$ defined in (\ref{csta}) which assumes two values, depending on  $\y$, we will analyze the two cases separately.

\ni
We start by  analyzing  (\ref{eqcondx}) when   $\y>-1$. In this case,  $h(\y)=d(1+\y)$ and so  $e^{2d\beta \xx} e^{\beta h(\yy)}=e^{\beta d[2\xx+(1+\yy)]}$. Since in the disordered region  $1+2\x+\y<0$, we have $\delta\leq 1$. So, \eqref{eqcondx} holds if  \begin{eqnarray}
\label{eqcondxz}-\log{(1-8de\delta\epsilon)}^{1\over 8de\delta\epsilon}\leq {5\over 4}.
\end{eqnarray}
Notice that the function $f(x)=-\ln[(1-x)^{1/x}]$ is increasing for $0<x<1$ with $\lim_{x\to 0^+}f(x)=1$ and $\lim_{x\to 1^-}f(x)=+\infty$.
The solution of  $f(x)=5/4$ is (slightly) greater that  $0.37137$. And so, the  condition (\ref{eqcondxz}) holds, provided that
$$\delta\epsilon\leq {0.37137\over 8de}\le{1\over 58.57d}.$$
Calling $-d[2\x+(1+\y)]=k_1$,  $1+|\y|=k_2$, and $58.57d=C_+$, the condition above becomes

\begin{equation*}
e^{-k_1 \beta }(1-e^{-k_2 \beta })\leq \frac{1}{C_+}.
\end{equation*}
Note that $k_2\ge 0$ by definition and $k_1\ge 0$ since we are in the disordered phase where $2\x+(1+\y)<0$.
 Observe moreover that the maximum of the function $g(\beta)=e^{-k_1 \beta }(1-e^{-k_2 \beta })$ as $\b$ varies in the interval $(0,+\infty)$ is attained at $\b=\b_c$ where $\b_c$ is the solution of   $e^{-k_2\beta}=\frac{k_1}{k_1+k_2}$. So, for any $\b>0$ we have

\begin{equation*}
g(\beta)\leq \left(e^{-k_2\beta_c}\right)^{\frac{k_1}{k_2}}(1-e^{-k_2\beta_c})
=\left(\frac{k_1}{k_1+k_2}\right)^{\frac{k_1}{k_2}}\frac{k_2}{k_1+k_2}%=\frac{1}{\left(1+\frac{1}{t}\right)^{k_1/k_2}}\frac{k_2}{k_1+k_2}
\leq \frac{k_2}{k_1+k_2},
\end{equation*}
 %since $1\leq (1+1/t)^t<e$.
 Therefore, if $\frac{k_2}{k_1+k_2}\leq \frac{1}{C_+} $, or equivalently, $k_1\geq (C_+-1)k_2$, the condition for convergence will be satisfied for every $\b\ge 0$. Taking into account the expressions for  $k_1$ and $k_2$, we have

\begin{eqnarray*}
-d[2 \x+(1+\y)]\geq (C_+-1)(1+|\y|),
\end{eqnarray*}
%which is equivalent to
%\begin{eqnarray*}
% 2d|x|\geq (C_+-1)(1+|y|)+d(1+y),
% \end{eqnarray*}
namely,
\begin{eqnarray*}
  \x\leq -\frac{(C_+-1)}{2d}(1+|\y|)-\frac{1}{2}(1+\y).
\end{eqnarray*}
Since we have dependency on $|\y|$, we have to consider the cases $\y\geq 0$ and  $-1<\y<0$ separately (remember that we are currently considering the case $\y>-1$). For $\y\geq 0$, we have
\begin{equation*}\label{caso1.1}
\x\leq -\left[\frac{(C_+-1)}{2d}+\frac{1}{2}\right](1+\y)
\end{equation*}
and for $-1<\y<0$,  we have

\begin{equation*}\label{caso1.2}
 \x\leq -\left[\frac{(C_+-1)}{2d}+\frac{1}{2}\right]+\left[\frac{(C_+-1)}{2d}-\frac{1}{2}\right] \y.
 \end{equation*}

\ni
Let us now turn to  the case when $\y\leq -1$. In this case the stability constant is $h(\y)=0$ and thus, recalling that
$\d=\alpha(\x,\b) e^{\b h(\yy)}$ and that $\alpha(\x,\b) =\frac{e^{2d\beta\,\xx}}{1+2e^{2d\beta \xx}}$ we have that    $\delta\leq {1\over 3}$. Therefore, the condition \eqref{eqcondx} becomes
$$
-\log{(1-8de\delta\epsilon)}^{1\over 8de\delta\epsilon}\leq {7\over 4},
$$
namely, $$\delta\epsilon\leq {0.7127\over 8de}\le {1\over 30.52d}.$$ Letting $k_1=2d|x|$, $k_2=1-y$ and $C_-=30.52d$ and  proceeding as  in the first case, we have the following condition

\begin{equation*}\label{caso2}
\x\leq -\frac{(C_--1)}{2d}(1-\y).
\end{equation*}
Summarizing, we conclude that the free energy is analytic for all $\beta$ for $(x,y)$ in the portion of ${\cal D}$ whose boundary is the polygonal curve
 \begin{eqnarray*}\x=\left\{\begin{array}{ll}
 -k(\y+1),& \mbox{if $\y\geq0$}\\
 (k-1)\y-k,&\mbox{if $-1<\y< 0$}\\
 \bar{k}(\y-1),&\mbox{if $\y\leq -1$}
 \end{array} \right.,
 \end{eqnarray*}
where  $k={59.57d-1\over 2d}$ and  $\bar{k}={30.52d-1\over 2d}$, and this concludes the proof of  Theorem \ref{teo1}.

\def\\{\noindent}\def\La{\Lambda}\def\Z{\mathbb{Z}}\def\zd{\mathbb{Z}^d}
\def\AA{{\mathcal A}}
\section{A remark on the numbers of $d$-dimensionl polycubes of size $n$  }\label{poly}
We conclude this note by showing that Lemma \ref{lemma2} proved  in the previous section
can be used to obtain an upper bound for the number $A_n$ of fixed $d$-dimensional polycubes of size $n$.
We recall that
a $d$-dimensional polycube of size $n$ is a connected set of $n$ unit cubical cells on the
lattice $\Zd$, where connectivity is through $(d-1)$-faces. Let us denote by $\PP_n$ the set of all $d$-dimensional polycubes of size $n$
in $\Zd$.
Two polycubes
are considered equivalent if one can be transformed into the other by a translation. A class of equivalence of polycubes
is called a ``fixed polycube". Let us denote by $A_n$ the number of fixed $d$-dimensional polycubes of size $n$ and let us
explain how obtain an upper bound for $A_n$.

\ni
First of all let us give the following definition.  A finite set of vertices $Q$ in $\mathbb{Z}^d$
is  called an {\it animal}  if either $|Q|= 1$ or if $|Q|\ge 2$ and for any partition $Q=A\uplus B$,  there exist $x\in A$ and $y\in B$ such that
$|x-y|=1$. Let us denote by $\AA_n$ the set of all animals in $\Zd$ with $n$ vertices. Clearly there is a one-to-one correspondence between $\PP_n$
and $\AA_n$. Indeed to each $Q=\{x_1,\dots,x_n\}\in \AA_n$ we can associate in a ono-to-one manner the polycube $p=\{c_1,\dots, c_n\}$ in the dual lattice $(\Zd)^*$, such that each cube
$c_i\in p$  is centered  in $x_i$.

\ni
Let us define
$$
A^*_n= \sum_{Q\in \AA_n:\atop 0\in Q} 1.
$$
Then, since there are $n$ possible choices of the position of the origin,
\be\label{aenne}
A_n={A^*_n\over n}
\ee
is the number
of fixed polyominoes (polycubes in three or more dimensions) of size $n$.
Let us consider the infinite graph $\mathbb{G}^d $ with set of vertices $\zd$ and set of edges formed by the nearest neighbor pairs of $\zd$.
Let us denote by $\mathbb{T}^d$ the set of all finite subgraphs of $\mathbb{G}^d $ which are trees.
Then  observe that any animal  $Q$ in $\zd$ contains at least one spanning tree in $\mathbb{T}^d$. Therefore,
$$
A^*_n=\sum_{Q\in \AA_n:\atop 0\in Q} 1
\le\sum_{\t\in \mathbb{T}^d: \atop 0\in V_\t, ~|V_\t|=n}1
$$
where $\t$ denotes a  spanning tree,  with $E_\tau$ and $V_\t$ its sets of edges and  vertices, respectively.

\\Now recall that we have denoted by  $T^*_n$  the set of all trees  with vertex set $[n]$ and degrees  $d_i$ such that $1\le d_i\le 2d$.
Recall also that, given two vertices $x_i,x_j$ in $\zd$,   $\d_{|x_i-x_j|1}=1$, if $x_i,x_j$ are neighbors in $\zd$ and $\d_{|x_i-x_j|1}=0$,  otherwise. Then we can write
$$
\sum_{\t\in \mathbb{T}^d: \atop 0\in V_\t, ~|V_\t|=n}1~ \le ~\sum_{\{x_1,\dots, x_n\}\subset \zd:\atop x_1=0}\sum_{\t\in T^*_n}
\prod_{\{i,j\}\in E_\t}\d_{|x_i-x_j|1}
$$
$$
~~~~~~~~~~~~~~~~~~~~~~~~\;~~~~=~{1\over (n-1)!}
\sum_{(x_1,\dots, x_n)\in \mathbb{Z}^{dn}:\atop x_1=0,~x_i\neq x_j}\sum_{\t\in T^*_n}
\prod_{\{i,j\}\in E_\t}\d_{|x_i-x_j|1}~
$$
$$
~~~~~~~~~~~~~~~~~~~~~\;~~~~=~{1\over (n-1)!}\sum_{\t\in T^*_n}
\sum_{(x_1,\dots, x_n)\in \mathbb{Z}^{dn}:\atop x_1=0,~x_i\neq x_j}
\prod_{\{i,j\}\in E_\t}\d_{|x_i-x_j|1}
$$
$$
=~{1\over (n-1)!}\sum_{\t\in T^*_n}
w_\t ~~~~~~
$$
where $w_\t$ is defined in (\ref{wtau0}).
Now given a tree $\t\in T^*_n$ with degree $d^\t_1,\dots d^\t_n$ at vertices $1,\dots ,n$, we can use the bound (\ref{wtau}) and
hence we get
$$
\sum_{\t\in \mathbb{T}^d: \atop 0\in V_\t, ~|V_\t|=n}1
\le {2d\over (n-1)!}\sum_{\t\in T^*_n}\prod_{i=1}^{n} {(2d-1)!\over (2d-d^\t_i)!}~~~~~~~~~~~~~~~~~~~~~~~~~~~~~~~~~~~~~
$$
$$
~~~~~= {2d\over (n-1)!} \sum_{d_1+\dots +d_n=2n-2\atop 1\le d_i\le 2d} \prod_{i=1}^{n} {(2d-1)!\over (2d-d_i)!}
\sum_{\t\in T^*_n\atop d_1,\dots, d_n~{\rm fixed}}1
$$
$$
~~~~~= {2d\over (n-1)!} \sum_{d_1+\dots +d_n=2n-2\atop 1\le d_i\le 2d} \prod_{i=1}^{n} {(2d-1)!\over (2d-d_i)!}
{(n-2)!\over \prod_{i=1}^n(d_i-1)}
$$
$$
= {2d\over (n-1)}\sum_{s_1+\dots +s_n=n-2\atop 0\le s_i\le 2d-1}
\prod_{1=1}^{n}{2d-1 \choose s_i}.~~~~~~~~~~~~~~
$$
Now observe that
$$
\sum_{s_1+\dots +s_n=n-2\atop 0\le s_i\le 2d-1}
\prod_{1=1}^{n}{2d-1 \choose s_i}= {(2d-1)n\choose n-2}.
$$
Indeed, suppose that we have a set $A=\biguplus_{i=1}^nA_i$ and each $A_i$  has cardinality $2d-1$.
Hence, for fixed numbers $s_1, \dots , s_n$ such that $0\le s_i\le 2d-1$, we have that $\prod_{1=1}^{n}{2d-1 \choose s_i}$
is the number of ways to choose $s_1$ objects from $A_1$, $s_2$ objects from $A_2$, $\dots,$, $s_n$ objects from $A_n$.
In this way we have chosen $s_1+\dots+s_n$ objects in $A$. Since $s_1+\dots+s_n=n-2$, this is the same as to choose
$n-2$ elements in $A$  which is $(2d-1)n\choose n-2$.
Therefore we have obtained the bound
$$
A_n^*\le  {2d\over (n-1)} {(2d-1)n\choose n-2}
$$
and consequently, recalling (\ref{aenne}),  an upper  bound for  the number $A_n$
of fixed $d$-dimensional polycubes of size $n$, namely,
\be\label{polycubes}
A_n\le  A^{\rm LLP}_n\doteq {2d\over n(n-1)} {(2d-1)n\choose n-2}.
\ee
The bound (\ref{polycubes}) is slightly   better than  previous bounds available in the literature, at least  for $d\ge 3$ (see e.g.
Theorem 9 in \cite{BBR} and   \cite{BS}). In particular,  in Section 2 of  the very recent paper \cite{BS}   Barequet and  Shalah claim that
$A_n$ is bounded from above by the number of binary sequences with
$n-1$ ones and $(2d-2)n$ zeros and hence
$$A_n\le A^{\rm BS}_n\doteq{(2d-1)n-1\choose n-1}$$
so that the ratio
$${A_n^{\rm LLP}\over A_n^{\rm BS}}={2d(2d-1)\over ((2d-2)n+1)(2d-2)n+2)}$$
goes to zero as $O(1/n^2)$ when $n\to \infty$.

\section{Conclusions}
In this paper we analyze the $d$-dimensional, with $d\geq 2$, Blume-Emery-Griffiths model (with Hamiltonian given by (\ref{hamiltoniana1})) in the disordered region of parameters ${\cal D}\equiv \{(\x,\y)\in\mathbb{R}^2:\x<0, \, 1+2\x+\y<0\}$
and we obtain a region ${\cal D}_{analytic}\subset {\cal D}$ where
the finite volume free energy of the model  can be written, at any nonnegative inverse temperature $\b$, in terms of  an absolutely convergent series,  uniformly bounded
in the volume. This implies that  its thermodynamic limit is an analytic function
of  $\b$, for all nonnegative $\b$, ruling out the possibility of a phase transition in the region ${\cal D}_{analytic}$.
This result has been obtained via a high temperature polymer expansion of the partition function of the model combined with  some recent results related to tree graph inequalities.  We believe that our estimates   are nearly optimal compatibly with the cluster expansion techniques used here. In confirmation of this, we also obtain,  as a  byproduct, a slight improvement on the upper bound for the number of $d$-dimensional fixed polycubes of size $n$.  On the other hand, we think that  to enlarge  sensibly the region of ${\cal D}_{analytic}$ in which no phase transition
occurs for all nonnegative $\beta$, in particular in such a way to include
negative values of the
parameter $\x$  near the line $\x=0$ and $\y<-1$, new ideas are necessary.

\section*{Acknowledgements} This work has been partially supported by the Brazilian agencies
Coordenadoria de Aperfei\c{c}oamento de Pessoal de N\'\i vel Superior (CAPES) and
 Conselho Nacional de Desenvolvimento Cient\'\i fico e Tecnol\'ogico (CNPq).

\end{document}